\documentclass[aps,prd,nofootinbib,floatfix,amssymb]{revtex4} 
\usepackage{epsf,feynmf}
\usepackage{graphicx,epsfig}
\usepackage{amsfonts}
\usepackage{amssymb}

\def\gsim{\mathrel {\vcenter {\baselineskip 0pt \kern 0pt \hbox{$>$} \kern 0pt \hbox{$\sim$} }}}

\def\lsim{\mathrel {\vcenter {\baselineskip 0pt \kern 0pt \hbox{$<$} \kern 0pt \hbox{$\sim$} }}}

\newcommand{\bdv}[1]{{\bf{#1}}}

\newcommand{\U}[2]{U_{#1}^{#2}}
\newcommand{\Ud}[2]{U_{#1}^{*}{}^{#2}}

\begin{document}

\title{The ``in-in'' Formalism and Cosmological Perturbations}
\author{Peter Adshead}
\author{Richard Easther}
\affiliation{Department of Physics\\ Yale University, New Haven, CT 06511 USA}
\author{Eugene A. Lim}
\affiliation{Department of Physics and ISCAP,  \\ Columbia University, New York, NY 10027 USA}

\begin{abstract}
We describe an efficient scheme for evaluating higher order contributions to primordial cosmological perturbations using the ``in-in'' formalism, which is the basis of modern calculations of non-Gaussian and higher order contributions to the primordial spectrum. We show that  diagrams with two or more vertices require careful handling. We present an implementation of the operator formalism in which these diagrams can be evaluated in  a simple and transparent fashion.    We illustrate our  methodology by evaluating the correction to the primordial gravitational wave spectrum generated by scalar loops, a 2-vertex, 1-loop interaction. We then look at a generalized $N$-point, 2-vertex diagram.
\end{abstract}

\maketitle
 
\section{Introduction}

The pairing of the Arnowitt-Deser-Misner (ADM) formulation of general relativity \cite{Arnowitt:1962hi} and the  ``in-in'' approach to quantum field theory \cite{Schwinger:1961,Bakshi:1962dv,Bakshi:1963bn, Keldysh:1964ud} is a remarkably powerful tool for analyzing cosmological perturbations\footnote{The formalism is the subject of two recent comprehensive monologues \cite{RammerBook,Hubook}, where the reader is referred to for further information.}. First applied to cosmology by Jordan \cite{Jordan:1986ug} and Calzetta and Hu \cite{Calzetta:1986ey}, Maldecena's treatment of primordial non-Gaussianities  \cite{Maldacena:2002vr} established it as the preferred approach to computing both $N$-point correlation terms \cite{Seery:2005wm, Seery:2005gb, Seery:2006js, Chen:2006xjb,
Seery:2006vu, Chen:2006nt, Chen:2008wn} and loop corrections \cite{Weinberg:2005vy, Sloth:2006nu, Sloth:2006az, Seery:2007wf, Seery:2007we, Adshead:2008gk,Dimastrogiovanni:2008af,Seery:2008ms} to the inflationary perturbation spectrum.  The perturbations generated by very simple models of inflation are accurately described by the long established lowest-order expressions \cite{Bardeen:1983qw,Starobinsky:1982ee,Guth:1982ec}. However, current \cite{Komatsu:2008hk} and forthcoming experiments \cite{Baumann:2008aq} will put tight limits on the primordial bispectrum (and possibly the trispectrum), and thus place tight constraints on inflationary models with  non-trivial perturbation spectra, so these calculations are of substantial practical importance.
 
The purpose of this  paper is to present an efficient and transparent scheme for implementing the operator formalism \cite{Weinberg:2005vy}. Additionally, we  warn that in some formulations, diagrams with more than one vertex\footnote{We are not restricting our attention to diagrams with a fixed number of external legs, so the order of a diagram here is simply the number of vertices, and not the number of loops -- the vertex-count is equivalent to the number of copies of the interaction Hamiltonian being correlated.} can become pathological if their external momenta or vertices are not distinguished from one another in some way.   The problem can be viewed as one of initial conditions -- ``in-in'' takes the initial states of some set of fields, $|{\rm in}\rangle$, and calculates the expectation value of some set of operators with respect to these fields at a later time $t$, $|\Omega(t)\rangle$.   Specifying the initial conditions typically amounts to choosing the  Bunch-Davies vacuum: the dynamical degrees of freedom behave like harmonic oscillators at very small length scales, and each mode is assumed to be in its ground state. There is then an operational \emph{prescription} that enforces this selection in field theoretic calculations.    As we show below,  \emph{specifying} the initial conditions implicitly restricts  the algebraic manipulations one may perform.   To calculate an expectation value at some arbitrary time in the interaction picture we need to evolve the in-state forward in time.   Operationally, the simplest way of incorporating this initial condition is to include some evolution in imaginary time, $ t \rightarrow t(1+i\epsilon)$.  There are (at least) two simple ways of expressing this choice. One may define the integration in the time evolution operator to run along a contour in the complex plane, rather than along the real axis. Alternatively one may analytically continue the time variable itself to include a small imaginary component, leaving the integral on the real axis. 
Importantly, some algebraic manipulations which are permitted in one formulation are forbidden in the other.   

Interestingly, this problem cannot arise in conventional ``in-out'' calculations, or in ``in-in'' calculations for diagrams with a single vertex -- this discussion is thus timely, since theoretical analyses of primordial perturbations have matured to the point where more complicated interactions are now routinely considered  \cite{Seery:2008ax,Gao:2009gd}.  We use the operator formalism introduced by Weinberg \cite{Weinberg:2005vy},  as it provides an efficient and transparent scheme for performing in-in calculations, and the specific approach we develop here  can significantly reduce the algebraic workload associated with a given diagram.

This paper is organized as follows. We review the operator formalism in Section \ref{sect:vacuum}, using the notation summarized in the Appendices of our previous paper \cite{Adshead:2008gk}. We review two different prescriptions for injecting the Bunch-Davies initial conditions, showing how to obtain consistent results within the formalism of Weinberg \cite{Weinberg:2005vy}. In the two following Sections we present sample calculations.  In \S \ref{sect:gravloops} we calculate scalar loop corrections to the graviton power spectrum. As one would expect, the scalar corrections to graviton (tensor) perturbations are too small to have any practical impact, but the calculation is a useful example of our overall methodology.  In \S \ref{sect:Npt} we write down a two vertex contribution to the  $N$-pt correlation function with $p$ internal lines.  With $N= 2$, $p=2$ this is a correction to the propagator with the same topology as the graviton loop correction of \S \ref{sect:gravloops},  and for $N=4$, $p=1$ we have the topology of the graviton \cite{Seery:2008ax} and scalar \cite{Gao:2009gd} exchange contributions to the scalar 4-point (trispectrum). 
We show that these diagrams can become problematic in limits where the external momenta are not distinguishable.  We demonstrate that explicitly injecting the initial conditions by deforming the time contours into the complex plane at the outset sidesteps any difficulties and substantially simplifies the algebra. We conclude in Section \ref{sect:conclusions}.

\section{Specifying Initial Conditions in  ``in-in''} \label{sect:vacuum}

In the in-in formalism, calculations of the expectation value, $\langle W(t)\rangle$, of a product of operators $W(t)$  at time $t$, require that we evaluate
\begin{equation} \label{eqn:ininformula}
\langle W(t)\rangle =\left\langle \left(T e^{-i\int_{-\infty}^{t}H_{\rm int}(t') dt'}\right)^{\dagger}  ~  W(t)~ \left(Te^{-i\int_{-\infty}^{t} H_{\rm int}(t'') dt''}\right)\right\rangle \, ,
\end{equation} 
where the fields on the right hand side are Heisenberg fields. The expectation value is taken with respect to the initial state, $|\rm in\rangle$, which we assume to be the Bunch-Davies vacuum. 
The interaction Hamiltonian  $H_{\mathrm{int}}$ is defined in the usual way, so that the total Hamiltonian $H$ is the  combination of $H_{\mathrm{int}}$  and  the  free-field Hamiltonian, $H_0$, 
\begin{equation}
H = H_0 + H_{\mathrm{int}}.
\end{equation}
Analyses of cosmological perturbations typically start with the Einstein-Hilbert action of general relativity together with the appropriate matter action. One then uses the ADM formulation  \cite{Arnowitt:1962hi} to obtain an action containing only dynamical degrees of freedom  \cite{Maldacena:2002vr, Seery:2007wf}. From the action one constructs the Hamiltonian by defining conjugate momenta, and separating out the quadratic from the higher order parts:
$H_0$ consists of terms that are quadratic in the perturbative degrees of freedom (and thus free), while $H_{\mathrm{int}}$ consists of all third and higher order terms \cite{Weinberg:2005vy}. The free Hamiltonian $H_{0}$ drives the evolution of the operators, while  $H_{\rm int}$ evolves the states. This separation is natural,  since in a homogenous and isotropic background we can find the eigenstates of the free field Hamiltonian at past infinity. 

The interaction terms generally have derivative couplings even when the action contains only canonical kinetic terms. These derivative couplings are the end result of perturbatively expanding the action. In models with non-standard kinetic terms such as DBI inflation \cite{Alishahiha:2004eh} or k-inflation \cite{ArmendarizPicon:1999rj}, derivative couplings are generically present at the outset. Consequently,  we must proceed carefully when canonically quantizing the theory. If $L_{\mathrm{int}}$ is the portion of the  \emph{action} with terms third order and higher, the usual expression for the interaction Hamiltonian $H_{\mathrm{int}} = -L_{\mathrm{int}}$ acquires corrections at fourth order, as  first shown in \cite{Adshead:2008gk}. In slow roll inflation these corrections are proportional to powers of the inflationary slow-roll parameters, but are generically unsuppressed in DBI or k-inflation  \cite{Gao:2009gd,Huang:2006eh}. In the more general case, for instance if one wants to calculate correlations in the radiation dominated era, there will be extra interaction terms.  

To make contact with the familiar ``in-out'' formalism of QFT, insert complete sets of states labeled by $\alpha$ and $\beta$ into equation (\ref{eqn:ininformula}), %
\begin{eqnarray}
\langle W(t)\rangle =\int d\alpha \int d\beta \langle 0 | \left(T e^{-i\int_{t_{0}}^{t}H_{\rm int}(t') dt'}\right)^{\dagger} |\alpha\rangle\langle\alpha|  W(t) |\beta\rangle\langle \beta| \left(Te^{-i\int_{t_{0}}^{t} H_{\rm int}(t'') dt''}\right)|0\rangle \, .
\end{eqnarray}
The interpretation is clear, the ``in-in'' correlation is the product of vacuum transition amplitudes (``in-out'') and a matrix element $\langle \alpha | W(t) |\beta\rangle$, summed over all possible ``out'' states. The ``in-in'' formalism is simply standard QFT, rigged to compute correlation functions at fixed time, given \emph{initial conditions} instead of asymptotic boundary conditions. Initial conditions in QFT are usually specified by finding the eigenstates of the free Hamiltonian $H_0$, and stipulating that the system begins in one (or some combination) of these eigenstates. If the system begins in the quantum mechanical vacuum,  this amounts to putting our system in the vacuum state of $H_0$ at the initial time. Operationally,  the vacuum is selected by redefining the range of $t$ to include a small imaginary component \cite{Weinberg:1995mt}, $t\rightarrow t+i\epsilon|t|$.

There are two (and possibly many more) ways one can incorporate this choice within a calculation:
\begin{enumerate}
\item{Define the time integration  in the time evolution operator to run over a contour in the complex plane:
\begin{equation}
Te^{-i\int_{-\infty}^t H_{\mathrm{int}}(t')dt'}\rightarrow Te^{-i\int_{-\infty(1+i\epsilon)}^t H_{\mathrm{int}}(t')dt'} \label{eqn:contourrotate}.
\end{equation}
With this choice Eqn. (\ref{eqn:ininformula}) becomes 
\begin{equation} \label{eqn:ininformula2}
\langle W(t)\rangle =\left\langle \left(T e^{-i\int_{-\infty(1+i\epsilon)}^{t}H_{\rm int}(t') dt'}\right)^{\dagger}  ~  W(t)~ \left(Te^{-i\int_{-\infty(1+i\epsilon')}^{t} H_{\rm int}(t'') dt''}\right)\right\rangle \, .
\end{equation} 
Once this is done, complex conjugating the time evolution operator means that the time-forward contour does not coincide with the time-backward contour.  
The vacuum specification has broken the time-symmetry of the forward and backward time integrals.  

}

\item{Analytically continue the interaction Hamiltonian occurring in the time evolution operator so that it becomes a function of a complex variable:
\begin{eqnarray}
H(t)\rightarrow H(t(1+i\epsilon)).
\end{eqnarray}
This is achieved  by analytically continuing each of the times occurring in the expansion of Eqn. (\ref{eqn:ininformula}). Since the momenta, $k$, generically appear with the conformal time $\tau$ in the combination $k\tau$ this procedure is equivalent to an analytic continuation of the momenta flowing through the vertex. In practice one needs only to analytically continue the times or momenta appearing in the exponents of the mode functions.}

\end{enumerate}

If we ignore this issue for a moment it is straightforward to  express Eqn. (\ref{eqn:ininformula})  as \cite{Weinberg:2005vy, Musso:2006pt};
\begin{eqnarray}\label{eqn:Wein2}
\langle W(t) \rangle = \sum_{N = 0}^{\infty}i^{N}\int_{-\infty}^{t}dt_{N}\int_{-\infty}^{t_{N}}dt_{N-1}...\int_{-\infty}^{t_{2}}dt_{1}\left\langle \left[H_{\rm int}(t_{1}), \left[H_{\rm int}(t_{2}), ... \left[H_{\rm int}(t_{N}), W(t)\right]...\right]\right]\right\rangle.
\end{eqnarray}
Eqn. (\ref{eqn:Wein2}) is formally consistent with Eqn. (\ref{eqn:ininformula}). However, this manipulation is only self-consistent if we use the second vacuum specification prescription  above. Recall that for any symmetric, holomorphic function, $f(t_{1}, t_{2}) = f(t_{2}, t_{1})$, 
\begin{equation} \label{eq:square}
\int_{a}^{b}dt_{1}\int_{a}^{b}dt_{2}f(t_{1}, t_{2}) = 2\int_{a}^{b}dt_{1}\int_{a}^{t_{1}}dt_{2}f(t_{1}, t_{2}).
\end{equation}
Because   $f$ is symmetric under the exchange of its arguments, the integral over the square region on the left hand  side of Eqn~\ref{eq:square} is  twice the right hand integral over the lower triangle, where $t_2<t_1$.  This result is easily generalized to more variables and moving from (\ref{eqn:ininformula}) to (\ref{eqn:Wein2}) requires repeated manipulations of this form. This manipulation requires that the integrations be interchangeable.  The vacuum specification in Method 1 above breaks the equivalence of the integrals arising from operators on the right and left of $W$ in Eqn. (\ref{eqn:ininformula2}).   Consequently, terms like $HWH$, together with the contour specification, prevent one consistently writing down Eqn. (\ref{eqn:Wein2}), as the lower triangle is no longer identical to the upper triangle  due to the asymmetry of the contour specification. Persisting with this approach risks losing information about part of the region of integration. In the second prescription, no contour is specified, so the manipulations above are perfectly safe.
 
However, in applying  Eqn. (\ref{eqn:Wein2}) one splits terms (e.g. $H(t_{1})W(t)H(t_{2})$) arising from Eqn. (\ref{eqn:ininformula}), which must be summed {\em before\/} any limits are taken, in order to avoid introducing unphysical divergences.  At second order, Eqn. (\ref{eqn:Wein2}) is
\begin{eqnarray}\nonumber
\langle W(t)\rangle_{2}& = &  -\bigg[\int_{-\infty}^{t}dt_{2}\int_{-\infty}^{t_{2}}dt_{1}\big\langle \big(H_{\rm int}(t_{1})H_{\rm int}(t_{2})W(t) +W(t)H_{\rm int}(t_{2})H_{\rm int}(t_{1})\\ && - H_{\rm int}(t_{1})W(t)H_{\rm int}(t_{2}) - H_{\rm int}(t_{2})W(t)H_{\rm int}(t_{1})\big)\big\rangle\bigg] \, .
\end{eqnarray}
The terms appear to occur in conjugate pairs, so it may appear that this expression  reduces to
\begin{eqnarray}\label{eqn:Wein2massaged}
\langle W(t)\rangle_{2}& = &  -2\Re\left[\int_{-\infty}^{t}dt_{2}\int_{-\infty}^{t_{2}}dt_{1}\left\langle H_{\rm int}(t_{1})H_{\rm int}(t_{2})W(t) - H_{\rm int}(t_{1})W(t)H_{\rm int}(t_{2})\right\rangle\right] \, .
\end{eqnarray}
This manipulation is valid for $HHW$ and $WHH$, but the vacuum prescription above prevents one from writing the $HWH$ terms in this fashion.

Finally, one might proceed from Eqn. (\ref{eqn:Wein2}) without employing either of the prescriptions above and regulate the oscillatory integrals in the far past by adding the appropriate small imaginary component to the initial point. In this case  Eqn. (\ref{eqn:Wein2massaged}) is again self-consistent and for many diagrams this approach will work without difficulty. The in-in formalism associates each vertex with a time integration, and provided each vertex has a distinct momentum flowing through it, these distinct momenta effectively keep track of the distinct regions of integration.   Adding all possible permutations of the momenta which arise from the sum over contractions -- as per Wick's theorem \cite{Wick:1950ee} -- sums over all tessellations of the restricted integration region and picks up all contributions.  This approach fails when the sum of the momenta at each vertex are not distinguished from one another -- as would be the case if one decides to make a specific choice about the external momenta before the time integrals are done -- in which case it can lead to spurious divergences,  as we will see below.  Consequently, our preferred approach is to work directly with Eqn.~(\ref{eqn:ininformula2}) rather than Eqn (\ref{eqn:Wein2}). This not only avoids unphysical separation of the terms, but involves much less algebra.

 \section{Gravitational Waves from Loops of Scalars} \label{sect:gravloops}

We begin with the one-loop correction to the graviton (tensor) power spectrum generated by loops of second order scalar modes. In spatially flat gauge, the degrees of freedom are scalar field fluctuations $\delta\phi$ and transverse-traceless metric fluctuations (gravitons) $\gamma_{ij}$.  At leading order in slow roll $\delta\phi$ couples to gravitons through the 3-point interaction \cite{Maldacena:2002vr} 
\begin{equation}
S = \frac{1}{2}\int d^{3}x d\tau a^{2}(\tau) \gamma^{ij}\partial_{i}\delta\phi\partial_{j}\delta\phi,
\end{equation}
where $\tau$ is the conformal time and the interaction $\delta_{ij}\gamma^{ij}\delta\phi\delta\phi $ vanishes by gauge choice.  This yields the interaction Hamiltonian, which, after moving to Fourier space, is
\begin{eqnarray}\label{eqn:Hint}
H_{\rm int}(\tau)  & = &  \frac{1}{2}k'^{i}k''^{j}\gamma_{ij}({\bf k}, \tau)\delta\phi({\bf k'}, \tau)\delta\phi({\bf k''},\tau).
\end{eqnarray}
Expanding the free fields in Fourier modes
\begin{eqnarray}\label{eqn:phiexpansion2}
\int \frac{d^{3}k}{(2\pi)^{3}}~\delta \phi_{\bf k} & = & \delta\phi(\bdv{x} , \tau ) = \int \frac{d^3 k}{(2\pi)^{3}}\,e^{i\bdv{k}\cdot\bdv{x}} \left[a(\bdv{k})\U{k}{}(\tau)+a^{\dagger}(-\bdv{k})\Ud{k}{}(\tau)\right], \\
\int \frac{d^{3}k}{(2\pi)^{3}}~ \gamma_{ij, {\bf k}}= \gamma_{ij}({\bf x}, \tau) & = & \sum_{s = +, \times}\int \frac{d^{3}k}{(2\pi)^{3}}e^{i{\bf k}\cdot{\bf x}}\left[b^{s}({\bf k})\epsilon^{s}_{ij}({\bf k})\gamma^{s}_{k}(\tau)+b^{\dagger s}(-{\bf k})\epsilon^{ s}_{ij}(-{\bf k})\gamma^{s *}_{k}(\tau)\right],
\end{eqnarray}
where the polarization tensors are normalized so that $\sum_{s,s'}\epsilon^{s}_{ij}
\epsilon^{ij,s'} = 2\delta^{ss'}$ and the mode functions, $U_{k}(\tau) $ and $\gamma_{k}(\tau) $, are given by the solutions to the free field equations of motion (obtained from the quadratic part of the action, see for example \cite{Mukhanov:1990me}.) In the de Sitter limit these are:
\begin{eqnarray} \label{eqn:modefunctions}
U_{k}(\tau) & = & \frac{H}{\sqrt{2 k^{3}}}(1-i k\tau)e^{ik\tau},\\
\gamma_{k}(\tau) & = & \frac{H}{\sqrt{2 k^{3}}}(1-ik\tau)e^{ik\tau}.
\end{eqnarray}
The propagators are then
\begin{eqnarray}
\langle\delta\phi_{\bf k}(\tau)\delta\phi_{\bf p}(\tau')\rangle  
& = & \U{k}{}(\tau)\Ud{p}{}(\tau')\delta^3(\bdv{k}+\bdv{p}), \label{eqn:wightman}\\
\langle \gamma^{s}_{ij, \bf k}(\tau)\gamma^{s'}_{lm, \bf p}(\tau')\rangle & = & \gamma_{ k}(\tau)\gamma^{*}_{ p}(\tau')\epsilon^{s}_{ij}({\bf k})\epsilon^{s'}_{lm}(-{\bf p})\delta_{ss'}\delta({\bf k}+{\bf p}).
\end{eqnarray}

To compute the one loop contribution to the tensor power spectrum, we evaluate\footnote{The 1-vertex 1-loop scalar loop contribution to the graviton 2-point correlator is scale-free, and  does not generate $\log$ corrections \cite{Adshead:2008gk,Dimastrogiovanni:2008af}.}
\begin{eqnarray}\nonumber\label{eqn:2pt2}
\int d^3x\; e^{i{\bf k}\cdot({\bf x}-\bf{x'})}\left\langle \gamma^{ij}(\tau_{*})\gamma_{ij}(\tau_{*}) \right\rangle & = & - 2\Re\bigg[ \int^{\tau_{*}}_{-\infty_{-}} d\tau_2 \int^{\tau_2}_{-\infty_{-}} d\tau_1 \langle H_{int}(\tau_1)H_{int}(\tau_2)\gamma^{ij}(\tau_{*})\gamma_{ij}(\tau_{*}) \rangle\bigg] \\ && +\int_{-\infty_{-}}^{\tau_{*}}d\tau_{1}\int_{-\infty_{+}}^{\tau_{*}}d\tau_{2}\langle H_{int}(\tau_1)\gamma^{ij}(\tau_{*})\gamma_{ij}(\tau_{*}) H_{int}(\tau_2)\rangle.
\end{eqnarray}
where $-\infty_{\pm} \equiv -\infty(1\pm i\epsilon)$ denotes the contour choice. the de Sitter limit a short calculation gives
\begin{eqnarray}\nonumber \label{eqn:bigloopint}
&\int d^3x\; e^{i{\bf k}\cdot({\bf x}-\bf{x'})}\left\langle \gamma^{ij}(\tau_{*})\gamma_{ij}(\tau_{*}) \right\rangle &=  -4(2\pi)^{6}H_{*}^{-4}\int\frac{d^{3}p'}{(2\pi)^{3}}\int\frac{d^{3}p''}{(2\pi)^{3}}\delta({\bf k}+{\bf p'}+{\bf p''})p'{}^{4}\sin^{4}(\theta)\\ \nonumber
&&\!\! \times \Bigg[ \Re\Big[\gamma^{*}_{ k}(\tau_{*})\gamma^{*}_{k}(\tau_{*})  \int_{-\infty_{-}}^{\tau_{*}}\frac{d\tau_{2}}{\tau_{2}^{2}}\gamma_{k}(\tau_{2})U^{*}_{\bf p'}(\tau_{2})U^{*}_{\bf p''}(\tau_{2})\int_{-\infty_{-}}^{\tau_{2}}\frac{d\tau_{1}}{\tau_{1}^{2}}\gamma_{k}(\tau_{1})U_{\bf p'}(\tau_{1})U_{\bf p''}(\tau_{1})\Big]\\
&&\!\! -\frac{1}{2}\gamma^{*}_{ k}(\tau_{*}) \gamma_{ k}(\tau_{*}) \int_{-\infty_{+}}^{\tau_{*}}\frac{d\tau_{2}}{\tau_{2}^{2}}\gamma^{*}_{ k}(\tau_{2})U^{*}_{ p'}(\tau_{2})U^{*}_{ p''}(\tau_{2})  \int_{-\infty_{-}}^{\tau_{*}}\frac{d\tau_{1}}{\tau_{1}^{2}}\gamma_{k}(\tau_{1})U_{ p'}(\tau_{1})U_{ p''}(\tau_{1})\Bigg],
\end{eqnarray}
where $\theta$ is the angle between the external momentum ${\bf k }$ and the momentum ${\bf p'}$. However, if we use (\ref{eqn:Wein2massaged}) to perform the calculation the third line above is replaced by (dropping irrelevant prefactors)
\begin{figure}
  \begin{fmffile}{gravloop}
  \unitlength =2mm
  \begin{fmfgraph*}(40,15)
  \fmfleft{in1}
  \fmfright{out1}
  \fmf{wiggly}{in1,v1}
  \fmf{wiggly}{v2,out1}
  \fmf{plain,left=0.5,tension=0.3}{v2,v1}
  \fmf{plain,left=0.5,tension=0.3}{v1,v2}
  \fmflabel{$\bf{k}$}{in1}
  \fmflabel{$\bf{k'}$}{out1}
  \end{fmfgraph*}
  \end{fmffile}
    \caption{The Feynman diagram corresponding to 2-point graviton correlation with a scalar loop.  As we explain below, the diagram may be evaluated off-shell, so the momenta on the external legs have distinct labels, to avoid unphysical divergences.}
\label{fig:gravloop}
    \end{figure}
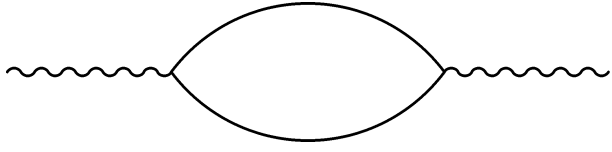
\begin{eqnarray}\nonumber\label{eqn:wronggravloop}
&& \gamma^{*}_{ k}(\tau_{*}) \gamma_{ k}(\tau_{*}) \Re\bigg[ \int_{-\infty}^{\tau_{*}}\frac{d\tau_{2}}{\tau_{2}^{2}}\gamma^{*}_{ k}(\tau_{2})U^{*}_{ p'}(\tau_{2})U^{*}_{ p''}(\tau_{2})  \int_{-\infty}^{\tau_{2}}\frac{d\tau_{1}}{\tau_{1}^{2}}\gamma_{k}(\tau_{1})U_{ p'}(\tau_{1})U_{ p''}(\tau_{1})\bigg]\\ 
& = & \frac{H^{6}}{8k^{3}p'^{3}p''^{3}}\Re\bigg[\gamma^{*}_{ k}(\tau_{*}) \gamma_{ k}(\tau_{*}) \int_{-\infty}^{\tau_{*}}\frac{d\tau_{2}}{\tau_{2}^{2}}(1+ia_{1}\tau_{2} - a_{2}\tau_{2}^{2}-ia_{3}\tau_{2}^{3})e^{-i a_{1}\tau_{2}}  \int_{-\infty}^{\tau_{2}}\frac{d\tau_{1}}{\tau_{1}^{2}}(1-ia_{1}\tau_{1} - a_{2}\tau_{1}^{2}+ia_{3}\tau_{1}^{3})e^{ia_{1}\tau_{1}}\bigg], \nonumber \\
\end{eqnarray} 
where $a_{1} = k+p'+p''$, $a_{2} = k(p''+p')+p'p''$ and $a_{3} = kp'p''$.  

In this case if we do not carefully track the vacuum specification we will always run into a divergence.  If one performs the integration on-shell,  the lower limit in the first integral may be made to vanish by adding a small amount of evolution in imaginary time \cite{Maldacena:2002vr}.  However, because the momentum flowing through each vertex is identical, the remaining integrand contains factors of $e^{a_1 \tau}$ and $e^{-a_1 \tau}$ which cancel, rendering  the second time integral divergent.  However, by writing down (\ref{eqn:wronggravloop}) we have implicitly ignored the difference in the contours between $\tau$ and $\tau'$. With the vacuum specification  explicitly included, the first integral leaves $-i (a_{1}(1+i\epsilon)-a_{1}(1+i\epsilon'))\tau$ in the exponential in the outer integral.\footnote{We are grateful to Xingang Chen  \cite{ChenNOTE}, David Seery, Martin Sloth and Filippo Vernizzi for discussions about this point.} Consequently as long as $\epsilon, \epsilon' \neq 0$ the integral is finite. The final limit $\epsilon, \epsilon' \rightarrow 0$ is rendered finite and order-independent by symmetrizing over $\epsilon, \epsilon'$, which is now required because  $H(t_{1})WH(t_{2})$ and $H(t_{2})WH(t_{1})$ are no longer conjugates. Performing this symmetrization is equivalent to performing the integration in the opposite order and thus picks up the other term.  Including the vacuum specification information in this fashion is equivalent to performing the calculation off shell and summing the result, before taking the on-shell limit.  Pursing this approach one obtains a term of form
\begin{eqnarray}\label{eqn:pain}
 \frac{H^{6}}{16k^{3}p'^{3}p''^{3}}
\gamma^{*}_{ k}(\tau_{*}) \gamma_{ k}(\tau_{*}) \left(\frac{a_{3}^2}{a_{1}^4}+\frac{2 a_{2} a_{3}}{a_{1}^3}-\frac{2
   a_{3}}{a_{1}}+\frac{a_{2}^2+a_{3}^2 \tau ^2}{a_{1}^2}+\frac{1}{\tau ^2}\right).
\end{eqnarray}

Much of this work can be avoided if one begins from Eqn. (\ref{eqn:ininformula2}), which yields Eqn. (\ref{eqn:bigloopint}). The integral factors into an expression of the form $|\int d\eta f(\eta)|^{2}$, which is not only well behaved everywhere but also reduces the double integral to a single integral. Carrying out the integration 
yields
\begin{eqnarray}
\frac{H^{6}}{16k^{3}p'^{3}p''^{3}}\gamma^{*}_{ k}(\tau_{*}) \gamma_{ k}(\tau_{*}) \left(-\frac{i (a_{1} a_{2}+a_{3})}{a_{1}^2}+\frac{a_{3} \tau }{a_{1}}-\frac{1}{\tau }\right) \left(\frac{i
   (a_{1} a_{2}+a_{3})}{a_{1}^2}+\frac{a_{3} \tau }{a_{1}}-\frac{1}{\tau }\right),
\end{eqnarray} 
which is quickly shown to be identical to the expression in Eqn. (\ref{eqn:pain}). The full expression is
\begin{eqnarray}\nonumber
\int d^3x\; e^{i{\bf k}\cdot({\bf x}-\bf{x'})}\left\langle \gamma^{ij}(\tau_{*})\gamma_{ij}(\tau_{*}) \right\rangle
& = & -\frac{(2\pi)^{6}}{2}\epsilon P_{\zeta\zeta}(k)P_{\gamma\gamma}(k) 
\int\frac{d^{3}p'}{(2\pi)^{3}}\int\frac{d^{3}p''}{(2\pi)^{3}}\frac{p'}{p''^{3}}\sin^{4}\theta{}\delta({\bf k}+{\bf p'}+{\bf p''})\\\nonumber && \Bigg[\frac{4 k^3+2 p'^2 k+5 p'^3}{ k}-\frac{\left(6 k^2+5 p'^2\right) p''}{
   k}+\frac{-6 k^4-5 p'^2 k^2-4 p'^3 k-5 p'^4}{ k K}\\\nonumber && -\frac{3
   \left(p'^4+2 k p'^3+3 k^2 p'^2+2 k^3 p'\right)}{ K^2} \\ &&-\left(k+p''-\frac{k^2+p''^2}{K}-\frac{p'' k^2+p''^2 k}{K^2}\right)^{2} +2\frac{ k p'' p''' }{K} - k^{2}\Bigg] ,
\end{eqnarray}
where $K = k+p'+p''$. The second and third lines come from the 1st term of Eqn. (\ref{eqn:bigloopint}) while the remaining terms come from the second term of Eqn. (\ref{eqn:bigloopint}). This expression takes the form
\begin{eqnarray}\label{eqn:lue}
I = -\frac{(2\pi)^{6}}{2}\epsilon P_{\zeta\zeta}(k)P_{\gamma\gamma}(k) \int d^3p' \sin^4 \theta \int d^3p'' \sum_{\alpha}\frac{f_{\alpha}(p',p'',k)}{K^{\alpha}},
\end{eqnarray}
where  $f_{\alpha}$ denote functions of $(p',p'',k)$ associated with the $\alpha$-th power of $K$.  We can eliminate the $\sin ^4 \theta$ term in the integral with the useful identity
\begin{eqnarray}\nonumber
&&\int\, d^{3}p\int\, d^{3}p'\delta^{3}({\bf p}+{\bf p'}+ {\bf
k})\sin^{2n}\theta f(p,p'k)=\\&&\quad\quad\quad (-1)^{n} \frac{2\pi}{k}\int_{0}^{\infty}\, p
dp\int_{|p-k|}^{p+k}\,p' d p'\left[\frac{K(2k-K)(K-2p')(K-2p'')}{4k^{2}p'^{2}}\right]^{n}f(p,p',k).
\end{eqnarray}
We are only interested in the $\log k$ contributions to this integral, and the $\sin^{4}\theta$ term effectively contains a factor $K^{2}$, so the only terms in equation (\ref{eqn:lue}) that are not simply polynomial divergences are those  with $\alpha>2$. After some straightforward algebra, we obtain
\begin{eqnarray}\nonumber \label{eqn:gravloopint}
\int d^3x\; e^{i{\bf k}\cdot({\bf x}-\bf{x'})}\left\langle \gamma^{ij}(\tau_{*})\gamma_{ij}(\tau_{*}) \right\rangle 
&=&\!\!\!\!\frac{\pi}{k}\epsilon P_{\zeta\zeta}(k)P_{\gamma\gamma}(k) \\ && \!\!\!\!\! \times\int_{0}^{\infty}\,  dp'\int_{|p'-k|}^{p'+k}\,d p'' \left[ 
\frac{p'^2p''^2}{K^2}-\frac{4p'{}^2p''k+5p'{}^3p''+5p''{}^2p'{}^2+p'p''{}^3}{kK} + ...\right],
\end{eqnarray}
where the ellipses indicate polynomial terms which we will drop.
Eqn. (\ref{eqn:gravloopint}) is IR finite, as the integrand contains no negative powers of $p'$ or $p''$. Following \cite{Weinberg:2005vy}, we dimensionally regularize this expression to obtain \begin{equation} \label{eqn:finalloop}
\langle\gamma^{ij}\gamma_{ij}\rangle = P_{\gamma\gamma}\left(1-\frac{35}{4}\pi\epsilon P_{\zeta\zeta}\log k\right),
\end{equation}
where $P_{\zeta\zeta} = H^2/(4M_p^2k^{3}\epsilon)$ and $P_{\gamma\gamma} = H^2/(M_p^2k^{3})$, which are the uncorrected primordial curvature and tensor power spectra respectively. The loop corrections to the tensor power spectrum via a scalar loop are $\sim (H/M_p)^4$, and thus minute, as we would expect.

\begin{figure}  \begin{fmffile}{Npoint}  \unitlength =1.5mm
\begin{fmfgraph*}(40,15)  \fmfleft{in1,in2,in3,in4}
\fmfright{out1,out2,out3,out4}  \fmf{plain}{in1,v1}
\fmf{plain}{in2,v1}  \fmf{plain}{in3,v1}  \fmf{plain}{in4,v1}
\fmf{plain}{v2,out1}
 \fmf{plain}{v2,out2}
 \fmf{plain}{v2,out3}  \fmf{plain}{v2,out4}
\fmf{plain,pull=5,left=0.9,tension=0.7}{v1,v2}
\fmf{plain,pull=5,left=0.5,tension=0.7}{v1,v2}
\fmf{plain,pull=5,left=-0.5,tension=0.7}{v1,v2}
\fmf{phantom,pull=5,left=-0.7,tension=0.7}{v1,v2}
\fmf{plain,pull=5,left=-0.9,tension=0.7}{v1,v2}  \fmflabel{${\bf
k}_1$}{in1}  \fmflabel{${\bf k}_2$}{in2}  \fmflabel{${\bf k}_3$}{in3}
\fmflabel{${\bf k}_4$}{in4}  \fmflabel{${\bf k}_5$}{out1}
\fmflabel{${\bf k}_6$}{out2}  \fmflabel{${\bf k}_7$}{out3}
\fmflabel{${\bf k}_8$}{out4}  \end{fmfgraph*}  \end{fmffile}
\caption{A sample diagram with $N=8$ legs, two $m=8$ point vertices
and $p=4$ internal lines.}\label{fig:npoint}    \end{figure}
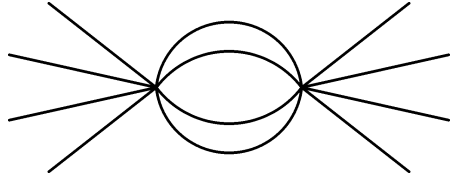

\section{The N-point 2-vertex Interaction} \label{sect:Npt}

The splitting of the terms in the expansion of (\ref{eqn:ininformula})  can induce spurious divergences in general multi-vertex topologies, as well as the 2-point diagram considered above. These divergences necessarily cancel in any careful calculation, but we now show that the problem can arise for a general $N$-point function.  We consider the class of diagrams with $N$ external legs, 2 vertices, and $p$ internal lines  -- an example is given in Figure~\ref{fig:npoint}.  

Suppose we have a term in the interaction Hamiltonian of the form $f(\tau)\delta\phi^{m}(\tau)$, where $f(\tau)$ is some coupling. At second order one generates a general $N$-point function through a 2-vertex process with $p$ internal lines. For simplicity we restrict ourselves to symmetric diagrams where each vertex has the same form, and  with no loops which begin and end on the same vertex, so there are $p= m-N/2$ internal lines and $N$ is even. At second order by this diagram is generated by
\begin{eqnarray}\label{eqn:2vertexN} \nonumber
\langle \delta \phi(\tau_{*})^{N}\rangle_{2}& = &  -2\Re\bigg[\int_{-\infty}^{\tau_{*}}d\tau_{2}f(\tau_{2})\int_{-\infty}^{\tau_{2}}d\tau_{1}f(\tau_{1})\big\langle \delta\phi^{m}_{\rm int}(\tau_{1})\delta\phi^{m}_{\rm int}(\tau_{2}) \delta\phi(\tau_{*})^{N}\rangle\Bigg] \\
&& + \int_{-\infty}^{\tau_{*}}d\tau_{2}f(\tau_{2})\int_{-\infty}^{\tau_{*}}d\tau_{1} f(\tau_{1})\langle \delta\phi^{m}_{\rm int}(\tau_{1})\delta\phi(\tau_{*})^{N}\delta\phi^{m}_{\rm int}(\tau_{2})\big\rangle\, .
\end{eqnarray}
If one had used the expansion of Eqn. (\ref{eqn:Wein2}), one would have
\begin{eqnarray}\nonumber\label{eqn:2vertexNwein}
\langle\delta \phi(\tau_{*})^{N}\rangle_{2}& = &   -2\Re\bigg[\int_{-\infty}^{\tau_{*}}d\tau_{2}f(\tau_{2})\int_{-\infty}^{\tau_{2}}d\tau_{1}f(\tau_{1})\big\langle \delta\phi^{m}_{\rm int}(\tau_{1})\delta\phi^{m}_{\rm int}(\tau_{2}) \delta\phi(\tau_{*})^{N}\rangle\Bigg] \\ && + \int_{-\infty}^{\tau_{*}}d\tau_{2}f(\tau_{2})\int_{-\infty}^{\tau_{2}}d\tau_{1} f(\tau_{1})\bigg(\langle \delta\phi^{m}_{\rm int}(\tau_{1})\delta\phi(\tau_{*})^{N}\delta\phi^{m}_{\rm int}(\tau_{2}) + \delta\phi^{m}_{\rm int}(\tau_{2})\delta\phi(\tau_{*})^{N}\delta\phi^{m}_{\rm int}(\tau_{1})\big\rangle\bigg) \, .
\end{eqnarray}
where the second term in Eqn. (\ref{eqn:2vertexN}) has been split into two pieces in Eqn. (\ref{eqn:2vertexNwein}). In the de Sitter limit the second term becomes
\begin{eqnarray}\nonumber\label{eqn:Npointdesitter}
 \langle \delta\phi^{m}_{\rm int}(\tau_{1})\delta\phi(\tau_{*})^{N}\delta\phi^{m}_{\rm int}(\tau_{2})\big\rangle & = & p!\prod_{i = 1}^{N/2}\frac{H^{2}}{2k_{i}^{3}}(1-i k_{i}\tau_{1})(1+i k_{i}\tau_{*})e^{ik_{i}(\tau_{1}-\tau_{*})}\prod_{j = N/2}^{N}\frac{H^{2}}{2k_{j}^{3}}(1- i k_{j} \tau_{*})(1+ i k_{j} \tau_{2})e^{ik_{i}(\tau_{*}-\tau_{2})}\\\nonumber
 & & \times \prod_{k}^{p}\int \frac{d^{3}q_{k}}{(2\pi)^{3}}\frac{H^{2}}{2q_{k}^{3}}(1-iq_{k}\tau_{1})(1+iq_{k}\tau_{2})e^{iq_{k}(\tau_{1}-\tau_{2})}\delta^{3}(\sum_{i = 1}^{N/2}{\bf k}_{i}-\sum_{k=1}^{p}{\bf q}_{k})\delta^{3}(\sum_{i = 1}^{N/2}{\bf k}_{i}+\sum_{k=1}^{p}{\bf q}_{k})\\
&& + (N!-1)\;\;\text{permutations} \, .
\end{eqnarray}
The product of delta functions ensures overall momentum conservation, the $p!$ counts the number of equivalent ways of contracting the internal lines, and the final sum over the permutations  counts contractions of the external lines into the vertices. At this stage, the only difference between  Eqn. (\ref{eqn:2vertexN}) and Eqn. (\ref{eqn:2vertexNwein}) is the limits of integration, and  that Eqn.(\ref{eqn:2vertexNwein}) requires that we add the same term with the order of integration reversed. The first term can be obtained from this expression  simply by flipping the signs of the $k_{i}$ where $i\in\{1, ..., N/2\}$. Dropping all the irrelevant factors, suppressing the momentum integral and putting in the integration over the times, 
\begin{eqnarray}\nonumber\label{eqn:2vertexNintwein}
 &&\int_{-\infty}^{\tau_{*}}f(\tau_{2}) d\tau_{2}R(\tau_{2})e^{-i(\sum_{i = N/2}^{N} k_{i}+\sum_{l} q_{l})\tau_{2}} \int_{-\infty}^{\tau_{a}}f(\tau_{1})d\tau_{1}Q(\tau_{1})e^{i(\sum_{i = 1}^{N/2}k_{i}+\sum_{k}q_{k})\tau_{1}}\\\
 & & \times \delta(\sum_{i = 1}^{N/2}{\bf k}_{i}-\sum_{k=1}^{p}{\bf q}_{k})\delta(\sum_{i = N/2}^{N}{\bf k}_{i}+\sum_{k=1}^{p}{\bf q}_{k}) + (N!-1)\;\;\text{permutations} ,
\end{eqnarray}
where $Q$ and $R$ are the polynomials $Q(\tau_{1}) = \prod_{i = 1}^{N/2}(1-i k_{i}\tau_{1})\prod_{k}^{p}(1-iq_{k}\tau_{1})$ and $R(\tau_{2}) = \prod_{j = N/2}^{N}(1+ i k_{j} \tau_{2})\prod_{l}^{p}(1+iq_{l}\tau_{2})$. The upper limit of the $\tau_{2}$ integral is labelled by $\tau_{a}$, and we will take it to be either $\tau_{a} = \tau_{*}$ or $\tau_{a} =\tau_{2}$.  If we begin with Eqn. (\ref{eqn:2vertexN}), $\tau_{a} = \tau_{*}$  and Eqn. (\ref{eqn:Npointdesitter}) is simply the product of two integrals. Moreover, after exchanging the momenta these integrals are simply complex conjugates of one other. To evaluate the term one must perform a single integral. 

Conversely, if we use Eqn (\ref{eqn:2vertexNwein}), $\tau_{a} = \tau_{2}$ and we encounter similar complications to those seen with the loop in the previous section. Performing the first integral in Eqn. (\ref{eqn:2vertexNintwein}) with $\tau_{a} = \tau_{2}$ leaves  
\begin{equation}\label{eqn:2vertexNintwein1}
 \int_{-\infty}^{\tau_{*}}f(\tau_{2})d\tau_{2}R(\tau_{2})S(\tau_{2})e^{i(\sum_{i = 1}^{N}k_{i}-\sum_{i = N/2}^{N} k_{i})\tau_{2}} \delta(\sum_{i = 1}^{N/2}{\bf k}_{i}-\sum_{k=1}^{p}{\bf q}_{k})\delta(\sum_{j = N/2}^{N}{\bf k}_{j}+\sum_{k=1}^{p}{\bf q}_{k}) + (N!-1)\text{permutations} ,
\end{equation} 
where $S(\tau_{2})$ is another polynomial. Now, if one restricts to $\sum_{i = 1}^{N}k_{i}\rightarrow \sum_{j = N/2}^{N} k_{j}$ before\footnote{Faced with a complicated diagram, it is always tempting to restrict attention to situations with a high degree of symmetry.  An obvious example is the equilateral limit where all the momenta are of equal magnitude, which is a common choice in calculations of the 3-point function.  
 } performing these integrals one will encounter divergences. These divergences are spurious as discussed above. Properly including the vacuum specification prevents the limit from vanishing. Once the vacuum information is included the limit is really $\sum_{i = 1}^{N/2}k_{i}(\epsilon-\epsilon')$. Adding all the terms together (including those with the integration order reversed, which we have not written down)  renders the result finite and order independent in the limit $\epsilon,\epsilon'\rightarrow 0$.

 Overall momentum conservation, $\delta(\sum_{i}^{N}{\bf k}_{i})$, enforces $\left|\sum_{i = 1}^{N}{\bf k}_{i}\right| = \left|\sum_{j = N/2}^{N}{\bf k}_{j}\right|$ but does not require $\sum_{i = 1}^{N}k_{i} = \sum_{j = N/2}^{N} k_{j}$. Consequently, 
  problems only arise if we render the external momenta indistinguishable before performing the integrals. For the special case where $N=2$  overall momentum conservation  already requires that the external momenta are identical, so this problem shows itself immediately when computing simple loop corrections to the propagator. Employing an explicit vacuum specification at the outset amounts to working off-shell by an infinitesimal amount,  and thus ensures that the calculation is divergence-free.  This problem can arise in any multivertex diagram and is not restricted to loop graphs -- to see this, simply set $p=1$ in our general topology, reducing it to an $N$-point 2-vertex tree diagram.\footnote{With $N=4$ and $p=1$ the topology matches the graviton \cite{Seery:2008ax} and \cite{Gao:2009gd} scalar exchange contribution to the scalar 4-point (trispectrum).  While the specific details of the interaction differ, the algebra is very similar to that of the diagram we consider here.} 
\section{Conclusions} \label{sect:conclusions}

We have explored the role of initial state selection in the operator formulation of the ``in-in'' approach to quantum field theory, demonstrating that this requires careful handling at second order and above. Issues arise from the  two \emph{different} time contours which select the vacuum  $|\text{in}\rangle$ and $\langle\text{in}|$ at early times, and thus similar issues are not encountered in conventional in-out computations. To perform calculations, one uses the time evolution operator, $U(\tau, \tau_{0})$ to evolve the ``in'' state forward in time   to $\tau_{*}$ at which the correlation is to be computed. We show that many complications are avoided  by keeping the initial state prescription explicit throughout the computation. Moreover, avoiding expansions which  unnecessarily  split terms into pieces  which are {\em individually\/} unphysical simplifies many calculations and sidesteps these vacuum issues entirely.

This issue was first encountered in 1-loop corrections to the scalar propagator \cite{Adshead:2008gk}, and we again consider a 1-loop calculation here. We calculate the correction to the primordial gravitational wave spectrum produced  by  scalar bremsstrahlung during slow roll inflation. As one would expect, this is tiny,  of order $\sim  \epsilon P_{\gamma\gamma}(k) P_{\zeta\zeta}(k)\ln(k)$ where $P_{\zeta\zeta}(k)$ and $P_{\gamma\gamma}(k)$ are the (tree level) spectra of primordial curvature perturbations and gravitational waves respectively, and $\epsilon$ is the usual slow roll factor. 
We show that one can avoid (potential) spurious divergences and messy algebra by working only with the physical terms arising from Eqn. (\ref{eqn:ininformula}) rather than the expansion in Eqn. (\ref{eqn:Wein2}) which subdivides some terms into pieces which, considered alone, are unphysical.

We then show that one can encounter similar problems with these terms in any diagram with two or more vertices, including tree-level expressions.  We demonstrate this by considering the $N$-point function generated by a diagram with 2-vertices and $p\geq1$ internal lines. Here, spurious divergences are not ubiquitous, but appear if one specializes to highly symmetric momentum configurations before performing the time integrations. This problem is avoided if the vacuum selection is explicitly and consistently imposed throughout the calculation, even in these specialized limits.  In addition, eschewing seemingly convenient splits simplifies the resulting algebra, as a double integral is replaced with the product of a single integral and its complex conjugate.  Finally, while our analysis involves only 2-vertex interactions, Eqn. (\ref{eqn:Wein2}) splits up terms at every order, so this issue can arise in any multivertex diagram. None of these divergences will be physical, and in all cases the problem can be ameliorated by imposing the contour choice at the beginning of the calculation.

While it may seem that many higher order diagrams in cosmological perturbation theory are of purely academic interest as they relate to intrinsically miniscule effects (for an example where higher order effects can be important, see \cite{Tolley:2008qv}), as the quality of astrophysical data and  the theoretical sophistication of very early universe models increases, non-leading contributions to cosmological perturbations will become increasingly important. 

\section{Acknowledgments} 	
We thank   Hu Bin, Hael Collins,  Xian Gao, Richard Holman, and Steven Weinberg and for useful correspondence and discussions. We are  grateful to Xingang Chen for sharing an unpublished note, and we wish to particularly  thank David Seery, Martin Sloth and Filippo Vernizzi for a   detailed correspondence on this topic.  RE is supported in part by the United States Department of Energy, grant DE-FG02-92ER-40704 and by an NSF Career Award PHY-0747868.  
This research was supported by grant RFP1-06-17 from The Foundational Questions Institute (fqxi.org). EAL thanks KITP(China) and the Chinese Academy of Sciences for their hospitality where some of this work was done.

\bibliography{VacuumSelect}

\begin{thebibliography}{41}
\expandafter\ifx\csname natexlab\endcsname\relax\def\natexlab#1{#1}\fi
\expandafter\ifx\csname bibnamefont\endcsname\relax
  \def\bibnamefont#1{#1}\fi
\expandafter\ifx\csname bibfnamefont\endcsname\relax
  \def\bibfnamefont#1{#1}\fi
\expandafter\ifx\csname citenamefont\endcsname\relax
  \def\citenamefont#1{#1}\fi
\expandafter\ifx\csname url\endcsname\relax
  \def\url#1{\texttt{#1}}\fi
\expandafter\ifx\csname urlprefix\endcsname\relax\def\urlprefix{URL }\fi
\providecommand{\bibinfo}[2]{#2}
\providecommand{\eprint}[2][]{\url{#2}}

\bibitem[{\citenamefont{Arnowitt et~al.}(1962)\citenamefont{Arnowitt, Deser,
  and Misner}}]{Arnowitt:1962hi}
\bibinfo{author}{\bibfnamefont{R.}~\bibnamefont{Arnowitt}},
  \bibinfo{author}{\bibfnamefont{S.}~\bibnamefont{Deser}}, \bibnamefont{and}
  \bibinfo{author}{\bibfnamefont{C.~W.} \bibnamefont{Misner}}
  (\bibinfo{year}{1962}), \eprint{gr-qc/0405109}.

\bibitem[{\citenamefont{Schwinger}(1961)}]{Schwinger:1961}
\bibinfo{author}{\bibfnamefont{J.~S.} \bibnamefont{Schwinger}},
  \bibinfo{journal}{Proc. Nat. Acad. Sci.} \textbf{\bibinfo{volume}{46}},
  \bibinfo{pages}{1401} (\bibinfo{year}{1961}).

\bibitem[{\citenamefont{Bakshi and
  Mahanthappa}(1963{\natexlab{a}})}]{Bakshi:1962dv}
\bibinfo{author}{\bibfnamefont{P.~M.} \bibnamefont{Bakshi}} \bibnamefont{and}
  \bibinfo{author}{\bibfnamefont{K.~T.} \bibnamefont{Mahanthappa}},
  \bibinfo{journal}{J. Math. Phys.} \textbf{\bibinfo{volume}{4}},
  \bibinfo{pages}{1} (\bibinfo{year}{1963}{\natexlab{a}}).

\bibitem[{\citenamefont{Bakshi and
  Mahanthappa}(1963{\natexlab{b}})}]{Bakshi:1963bn}
\bibinfo{author}{\bibfnamefont{P.~M.} \bibnamefont{Bakshi}} \bibnamefont{and}
  \bibinfo{author}{\bibfnamefont{K.~T.} \bibnamefont{Mahanthappa}},
  \bibinfo{journal}{J. Math. Phys.} \textbf{\bibinfo{volume}{4}},
  \bibinfo{pages}{12} (\bibinfo{year}{1963}{\natexlab{b}}).

\bibitem[{\citenamefont{Keldysh}(1964)}]{Keldysh:1964ud}
\bibinfo{author}{\bibfnamefont{L.~V.} \bibnamefont{Keldysh}},
  \bibinfo{journal}{Zh. Eksp. Teor. Fiz.} \textbf{\bibinfo{volume}{47}},
  \bibinfo{pages}{1515} (\bibinfo{year}{1964}).

\bibitem[{\citenamefont{Rammer}(2007)}]{RammerBook}
\bibinfo{author}{\bibfnamefont{J.}~\bibnamefont{Rammer}},
  \emph{\bibinfo{title}{{Quantum Field Theory of Non-equilibrium States}}}
  (\bibinfo{publisher}{{C}ambridge, UK: Univ. Pr.}, \bibinfo{year}{2007}).

\bibitem[{\citenamefont{Hu and Calzetta}(2008)}]{Hubook}
\bibinfo{author}{\bibfnamefont{B.-L.} \bibnamefont{Hu}} \bibnamefont{and}
  \bibinfo{author}{\bibfnamefont{E.}~\bibnamefont{Calzetta}},
  \emph{\bibinfo{title}{{Nonequilibrium Quantum Field Theory}}}
  (\bibinfo{publisher}{{C}ambridge, UK: Univ. Pr.}, \bibinfo{year}{2008}).

\bibitem[{\citenamefont{Jordan}(1986)}]{Jordan:1986ug}
\bibinfo{author}{\bibfnamefont{R.~D.} \bibnamefont{Jordan}},
  \bibinfo{journal}{Phys. Rev.} \textbf{\bibinfo{volume}{D33}},
  \bibinfo{pages}{444} (\bibinfo{year}{1986}).

\bibitem[{\citenamefont{Calzetta and Hu}(1987)}]{Calzetta:1986ey}
\bibinfo{author}{\bibfnamefont{E.}~\bibnamefont{Calzetta}} \bibnamefont{and}
  \bibinfo{author}{\bibfnamefont{B.~L.} \bibnamefont{Hu}},
  \bibinfo{journal}{Phys. Rev.} \textbf{\bibinfo{volume}{D35}},
  \bibinfo{pages}{495} (\bibinfo{year}{1987}).

\bibitem[{\citenamefont{Maldacena}(2003)}]{Maldacena:2002vr}
\bibinfo{author}{\bibfnamefont{J.~M.} \bibnamefont{Maldacena}},
  \bibinfo{journal}{JHEP} \textbf{\bibinfo{volume}{05}}, \bibinfo{pages}{013}
  (\bibinfo{year}{2003}), \eprint{astro-ph/0210603}.

\bibitem[{\citenamefont{Seery and Lidsey}(2005{\natexlab{a}})}]{Seery:2005wm}
\bibinfo{author}{\bibfnamefont{D.}~\bibnamefont{Seery}} \bibnamefont{and}
  \bibinfo{author}{\bibfnamefont{J.~E.} \bibnamefont{Lidsey}},
  \bibinfo{journal}{JCAP} \textbf{\bibinfo{volume}{0506}}, \bibinfo{pages}{003}
  (\bibinfo{year}{2005}{\natexlab{a}}), \eprint{astro-ph/0503692}.

\bibitem[{\citenamefont{Seery and Lidsey}(2005{\natexlab{b}})}]{Seery:2005gb}
\bibinfo{author}{\bibfnamefont{D.}~\bibnamefont{Seery}} \bibnamefont{and}
  \bibinfo{author}{\bibfnamefont{J.~E.} \bibnamefont{Lidsey}},
  \bibinfo{journal}{JCAP} \textbf{\bibinfo{volume}{0509}}, \bibinfo{pages}{011}
  (\bibinfo{year}{2005}{\natexlab{b}}), \eprint{astro-ph/0506056}.

\bibitem[{\citenamefont{Seery and Lidsey}(2007)}]{Seery:2006js}
\bibinfo{author}{\bibfnamefont{D.}~\bibnamefont{Seery}} \bibnamefont{and}
  \bibinfo{author}{\bibfnamefont{J.~E.} \bibnamefont{Lidsey}},
  \bibinfo{journal}{JCAP} \textbf{\bibinfo{volume}{0701}}, \bibinfo{pages}{008}
  (\bibinfo{year}{2007}), \eprint{astro-ph/0611034}.

\bibitem[{\citenamefont{Chen et~al.}(2007{\natexlab{a}})\citenamefont{Chen,
  Easther, and Lim}}]{Chen:2006xjb}
\bibinfo{author}{\bibfnamefont{X.}~\bibnamefont{Chen}},
  \bibinfo{author}{\bibfnamefont{R.}~\bibnamefont{Easther}}, \bibnamefont{and}
  \bibinfo{author}{\bibfnamefont{E.~A.} \bibnamefont{Lim}},
  \bibinfo{journal}{JCAP} \textbf{\bibinfo{volume}{0706}}, \bibinfo{pages}{023}
  (\bibinfo{year}{2007}{\natexlab{a}}), \eprint{astro-ph/0611645}.

\bibitem[{\citenamefont{Seery et~al.}(2007)\citenamefont{Seery, Lidsey, and
  Sloth}}]{Seery:2006vu}
\bibinfo{author}{\bibfnamefont{D.}~\bibnamefont{Seery}},
  \bibinfo{author}{\bibfnamefont{J.~E.} \bibnamefont{Lidsey}},
  \bibnamefont{and} \bibinfo{author}{\bibfnamefont{M.~S.} \bibnamefont{Sloth}},
  \bibinfo{journal}{JCAP} \textbf{\bibinfo{volume}{0701}}, \bibinfo{pages}{027}
  (\bibinfo{year}{2007}), \eprint{astro-ph/0610210}.

\bibitem[{\citenamefont{Chen et~al.}(2007{\natexlab{b}})\citenamefont{Chen,
  Huang, Kachru, and Shiu}}]{Chen:2006nt}
\bibinfo{author}{\bibfnamefont{X.}~\bibnamefont{Chen}},
  \bibinfo{author}{\bibfnamefont{M.-x.} \bibnamefont{Huang}},
  \bibinfo{author}{\bibfnamefont{S.}~\bibnamefont{Kachru}}, \bibnamefont{and}
  \bibinfo{author}{\bibfnamefont{G.}~\bibnamefont{Shiu}},
  \bibinfo{journal}{JCAP} \textbf{\bibinfo{volume}{0701}}, \bibinfo{pages}{002}
  (\bibinfo{year}{2007}{\natexlab{b}}), \eprint{hep-th/0605045}.

\bibitem[{\citenamefont{Chen et~al.}(2008)\citenamefont{Chen, Easther, and
  Lim}}]{Chen:2008wn}
\bibinfo{author}{\bibfnamefont{X.}~\bibnamefont{Chen}},
  \bibinfo{author}{\bibfnamefont{R.}~\bibnamefont{Easther}}, \bibnamefont{and}
  \bibinfo{author}{\bibfnamefont{E.~A.} \bibnamefont{Lim}},
  \bibinfo{journal}{JCAP} \textbf{\bibinfo{volume}{0804}}, \bibinfo{pages}{010}
  (\bibinfo{year}{2008}), \eprint{0801.3295}.

\bibitem[{\citenamefont{Weinberg}(2005)}]{Weinberg:2005vy}
\bibinfo{author}{\bibfnamefont{S.}~\bibnamefont{Weinberg}},
  \bibinfo{journal}{Phys. Rev.} \textbf{\bibinfo{volume}{D72}},
  \bibinfo{pages}{043514} (\bibinfo{year}{2005}), \eprint{hep-th/0506236}.

\bibitem[{\citenamefont{Sloth}(2007)}]{Sloth:2006nu}
\bibinfo{author}{\bibfnamefont{M.~S.} \bibnamefont{Sloth}},
  \bibinfo{journal}{Nucl. Phys.} \textbf{\bibinfo{volume}{B775}},
  \bibinfo{pages}{78} (\bibinfo{year}{2007}), \eprint{hep-th/0612138}.

\bibitem[{\citenamefont{Sloth}(2006)}]{Sloth:2006az}
\bibinfo{author}{\bibfnamefont{M.~S.} \bibnamefont{Sloth}},
  \bibinfo{journal}{Nucl. Phys.} \textbf{\bibinfo{volume}{B748}},
  \bibinfo{pages}{149} (\bibinfo{year}{2006}), \eprint{astro-ph/0604488}.

\bibitem[{\citenamefont{Seery}(2008{\natexlab{a}})}]{Seery:2007wf}
\bibinfo{author}{\bibfnamefont{D.}~\bibnamefont{Seery}},
  \bibinfo{journal}{JCAP} \textbf{\bibinfo{volume}{0802}}, \bibinfo{pages}{006}
  (\bibinfo{year}{2008}{\natexlab{a}}), \eprint{0707.3378}.

\bibitem[{\citenamefont{Seery}(2007)}]{Seery:2007we}
\bibinfo{author}{\bibfnamefont{D.}~\bibnamefont{Seery}},
  \bibinfo{journal}{JCAP} \textbf{\bibinfo{volume}{0711}}, \bibinfo{pages}{025}
  (\bibinfo{year}{2007}), \eprint{0707.3377}.

\bibitem[{\citenamefont{Adshead et~al.}(2008)\citenamefont{Adshead, Easther,
  and Lim}}]{Adshead:2008gk}
\bibinfo{author}{\bibfnamefont{P.}~\bibnamefont{Adshead}},
  \bibinfo{author}{\bibfnamefont{R.}~\bibnamefont{Easther}}, \bibnamefont{and}
  \bibinfo{author}{\bibfnamefont{E.~A.} \bibnamefont{Lim}}
  (\bibinfo{year}{2008}), \eprint{0809.4008}.

\bibitem[{\citenamefont{Dimastrogiovanni and
  Bartolo}(2008)}]{Dimastrogiovanni:2008af}
\bibinfo{author}{\bibfnamefont{E.}~\bibnamefont{Dimastrogiovanni}}
  \bibnamefont{and} \bibinfo{author}{\bibfnamefont{N.}~\bibnamefont{Bartolo}},
  \bibinfo{journal}{JCAP} \textbf{\bibinfo{volume}{0811}}, \bibinfo{pages}{016}
  (\bibinfo{year}{2008}), \eprint{0807.2790}.

\bibitem[{\citenamefont{Seery}(2008{\natexlab{b}})}]{Seery:2008ms}
\bibinfo{author}{\bibfnamefont{D.}~\bibnamefont{Seery}}
  (\bibinfo{year}{2008}{\natexlab{b}}), \eprint{0810.1617}.

\bibitem[{\citenamefont{Bardeen et~al.}(1983)\citenamefont{Bardeen, Steinhardt,
  and Turner}}]{Bardeen:1983qw}
\bibinfo{author}{\bibfnamefont{J.~M.} \bibnamefont{Bardeen}},
  \bibinfo{author}{\bibfnamefont{P.~J.} \bibnamefont{Steinhardt}},
  \bibnamefont{and} \bibinfo{author}{\bibfnamefont{M.~S.}
  \bibnamefont{Turner}}, \bibinfo{journal}{Phys. Rev.}
  \textbf{\bibinfo{volume}{D28}}, \bibinfo{pages}{679} (\bibinfo{year}{1983}).

\bibitem[{\citenamefont{Starobinsky}(1982)}]{Starobinsky:1982ee}
\bibinfo{author}{\bibfnamefont{A.~A.} \bibnamefont{Starobinsky}},
  \bibinfo{journal}{Phys. Lett.} \textbf{\bibinfo{volume}{B117}},
  \bibinfo{pages}{175} (\bibinfo{year}{1982}).

\bibitem[{\citenamefont{Guth and Pi}(1982)}]{Guth:1982ec}
\bibinfo{author}{\bibfnamefont{A.~H.} \bibnamefont{Guth}} \bibnamefont{and}
  \bibinfo{author}{\bibfnamefont{S.~Y.} \bibnamefont{Pi}},
  \bibinfo{journal}{Phys. Rev. Lett.} \textbf{\bibinfo{volume}{49}},
  \bibinfo{pages}{1110} (\bibinfo{year}{1982}).

\bibitem[{\citenamefont{Komatsu et~al.}(2008)}]{Komatsu:2008hk}
\bibinfo{author}{\bibfnamefont{E.}~\bibnamefont{Komatsu}} \bibnamefont{et~al.}
  (\bibinfo{collaboration}{WMAP}) (\bibinfo{year}{2008}), \eprint{0803.0547}.

\bibitem[{\citenamefont{Baumann et~al.}(2008)}]{Baumann:2008aq}
\bibinfo{author}{\bibfnamefont{D.}~\bibnamefont{Baumann}} \bibnamefont{et~al.}
  (\bibinfo{year}{2008}), \eprint{0811.3919}.

\bibitem[{\citenamefont{Seery et~al.}(2008)\citenamefont{Seery, Sloth, and
  Vernizzi}}]{Seery:2008ax}
\bibinfo{author}{\bibfnamefont{D.}~\bibnamefont{Seery}},
  \bibinfo{author}{\bibfnamefont{M.~S.} \bibnamefont{Sloth}}, \bibnamefont{and}
  \bibinfo{author}{\bibfnamefont{F.}~\bibnamefont{Vernizzi}}
  (\bibinfo{year}{2008}), \eprint{0811.3934}.

\bibitem[{\citenamefont{Gao and Hu}(2009)}]{Gao:2009gd}
\bibinfo{author}{\bibfnamefont{X.}~\bibnamefont{Gao}} \bibnamefont{and}
  \bibinfo{author}{\bibfnamefont{B.}~\bibnamefont{Hu}} (\bibinfo{year}{2009}),
  \eprint{0903.1920}.

\bibitem[{\citenamefont{Alishahiha et~al.}(2004)\citenamefont{Alishahiha,
  Silverstein, and Tong}}]{Alishahiha:2004eh}
\bibinfo{author}{\bibfnamefont{M.}~\bibnamefont{Alishahiha}},
  \bibinfo{author}{\bibfnamefont{E.}~\bibnamefont{Silverstein}},
  \bibnamefont{and} \bibinfo{author}{\bibfnamefont{D.}~\bibnamefont{Tong}},
  \bibinfo{journal}{Phys. Rev.} \textbf{\bibinfo{volume}{D70}},
  \bibinfo{pages}{123505} (\bibinfo{year}{2004}), \eprint{hep-th/0404084}.

\bibitem[{\citenamefont{Armendariz-Picon
  et~al.}(1999)\citenamefont{Armendariz-Picon, Damour, and
  Mukhanov}}]{ArmendarizPicon:1999rj}
\bibinfo{author}{\bibfnamefont{C.}~\bibnamefont{Armendariz-Picon}},
  \bibinfo{author}{\bibfnamefont{T.}~\bibnamefont{Damour}}, \bibnamefont{and}
  \bibinfo{author}{\bibfnamefont{V.~F.} \bibnamefont{Mukhanov}},
  \bibinfo{journal}{Phys. Lett.} \textbf{\bibinfo{volume}{B458}},
  \bibinfo{pages}{209} (\bibinfo{year}{1999}), \eprint{hep-th/9904075}.

\bibitem[{\citenamefont{Chen et~al.}(2006)\citenamefont{Chen, Huang, and
  Shiu}}]{Huang:2006eh}
\bibinfo{author}{\bibfnamefont{X.}~\bibnamefont{Chen}},
  \bibinfo{author}{\bibfnamefont{M.-x.} \bibnamefont{Huang}}, \bibnamefont{and}
  \bibinfo{author}{\bibfnamefont{G.}~\bibnamefont{Shiu}},
  \bibinfo{journal}{Phys. Rev.} \textbf{\bibinfo{volume}{D74}},
  \bibinfo{pages}{121301} (\bibinfo{year}{2006}), \eprint{hep-th/0610235}.

\bibitem[{\citenamefont{Weinberg}(1995)}]{Weinberg:1995mt}
\bibinfo{author}{\bibfnamefont{S.}~\bibnamefont{Weinberg}},
  \emph{\bibinfo{title}{{The Quantum theory of fields. Vol. 1: Foundations}}}
  (\bibinfo{publisher}{{C}ambridge, UK: Univ. Pr.}, \bibinfo{year}{1995}).

\bibitem[{\citenamefont{Musso}(2006)}]{Musso:2006pt}
\bibinfo{author}{\bibfnamefont{M.}~\bibnamefont{Musso}} (\bibinfo{year}{2006}),
  \eprint{hep-th/0611258}.

\bibitem[{\citenamefont{Wick}(1950)}]{Wick:1950ee}
\bibinfo{author}{\bibfnamefont{G.~C.} \bibnamefont{Wick}},
  \bibinfo{journal}{Phys. Rev.} \textbf{\bibinfo{volume}{80}},
  \bibinfo{pages}{268} (\bibinfo{year}{1950}).

\bibitem[{\citenamefont{Mukhanov et~al.}(1992)\citenamefont{Mukhanov, Feldman,
  and Brandenberger}}]{Mukhanov:1990me}
\bibinfo{author}{\bibfnamefont{V.~F.} \bibnamefont{Mukhanov}},
  \bibinfo{author}{\bibfnamefont{H.~A.} \bibnamefont{Feldman}},
  \bibnamefont{and} \bibinfo{author}{\bibfnamefont{R.~H.}
  \bibnamefont{Brandenberger}}, \bibinfo{journal}{Phys. Rept.}
  \textbf{\bibinfo{volume}{215}}, \bibinfo{pages}{203} (\bibinfo{year}{1992}).

\bibitem[{\citenamefont{Chen}()}]{ChenNOTE}
\bibinfo{author}{\bibfnamefont{X.}~\bibnamefont{Chen}},
  \emph{\bibinfo{title}{Unpublished note}}.

\bibitem[{\citenamefont{Tolley and Wyman}(2008)}]{Tolley:2008qv}
\bibinfo{author}{\bibfnamefont{A.~J.} \bibnamefont{Tolley}} \bibnamefont{and}
  \bibinfo{author}{\bibfnamefont{M.}~\bibnamefont{Wyman}}
  (\bibinfo{year}{2008}), \eprint{0809.1100}.

\end{thebibliography}

\end{document}